**Biominerals with Texture Gradients are Functionally Graded Bioceramics Toughened by Stress Delocalization**

*David Wallis, Joe Harris, Corinna F. Böhm, Di Wang, Pablo Zavattieri, Patrick Feldner, Benoit Merle, Vitaliy Pipich, Katrin Hurle, Simon Leupold, Lars N. Hansen, Frédéric Marin, and Stephan E. Wolf\**

Dr. D. Wallis
Department of Earth Sciences, University of Cambridge, Cambridge, CB2 3EQ, United Kingdom.

Dr. J. Harris, Dr. Corinna F. Böhm, Simon Leupold, Dr. Stephan E. Wolf
Department of Materials Science and Engineering, Institute of Glass and Ceramics (WW3), Friedrich-Alexander University Erlangen-Nürnberg (FAU), Martensstrasse 5, 91058 Erlangen, Germany.
E-mail: stephan.e.wolf@fau.de

D. Wang, Prof. P. Zavattieri
Lyles School of Civil Engineering, Purdue University, West Lafayette, IN 47907-2051, Indiana, USA

P. Feldner, Dr. B. Merle
Department of Materials Science and Engineering, Institute I and Interdisciplinary Center for Nanostructured Films (IZNF), Friedrich-Alexander University Erlangen-Nürnberg (FAU), Cauerstr. 3, 91058 Erlangen, Germany.

Dr. V. Pipich
Jülich Centre for Neutron Science JCNS, Forschungszentrum Jülich GmbH,
Outstation at MLZ, Lichtenbergstr.1, 85747 Garching, Germany.

Dr. K. Hurle
GeoZentrum Nordbayern – Mineralogy, Friedrich-Alexander University Erlangen-Nürnberg (FAU), Schlossgarten 5a, 91054 Erlangen, Germany.

Prof. L. N. Hansen
Department of Earth and Environmnental Sciences, University of Minnesota, Minneapolis, 55455, United States of America.

Dr. F. Marin
UMR CNRS 6282 Biogéosciences, Université de Bourgogne Franche-Comté, 6 Boulevard Gabriel, 21000 Dijon, France

Dr. S. E. Wolf
Interdisciplinary Center for Functional Particle Systems (FPS), Friedrich-Alexander University Erlangen-Nürnberg (FAU), Haberstrasse 9a, 91058 Erlangen, Germany

Keywords: biominerals, gradient materials, bioinspired materials, ceramics.

**Abstract**

Biomineralizing organisms are widely noted and extensively studied due to their ability to generate structures exhibiting exceptional crystallographic control. Primarily, it is the organisms, such as sea-urchins or bivalves, that generate nearly single-crystalline biocrystals that have attracted attention. In contrast, biomineralizing organisms with seemingly disordered polycrystalline bio-armor have been left relatively unstudied. However, the crystalline ordering in the black-lipped pearl oyster, *Pinctada margaritifera*, reveals that biominerals with varying crystal textures are an unrecognized class of functionally graded materials. Changing crystal textures inevitably cause a variation in Young's modulus due to the orientation-dependent mechanical properties of crystals. The case of *Pinctada margaritifera* demonstrates that bioceramics with such crystallographical gradients are toughened by stress delocalization and reduced stress intensity factors, outperforming non-graded counterparts. These findings suggest that a multitude of biominerals, which are perceived as poorly ordered because of their polycrystallinity and changing crystal textures, may be considered as graded materials with hitherto unidentified emergent mechanical properties. The underlying design principle is remarkably simple and applicable to a wide range of crystalline material classes and may thus serve as a blueprint for future bioinspired functional materials.

**1. Introduction**

Biominerals are biogenic ceramic materials vital to the host organism's survival as they perform critical functions, such as sensors,[1,2] armor,[3] or weaponry.[4,5] Millions of years of selection pressure have optimized the designs of biominerals to the extent that evolutionarily-conserved structural motifs serve today as a library of inspiration for the design of solid-state materials.[6] As structural materials, the design of biominerals aims at attaining both strength and toughness, a combination of two materials traits that are typically mutually exclusive.[7]

The evolutionary optimization of these bioceramics thus addresses a classic material-design challenge; how to generate hard but non-brittle materials. Nature's solution to this challenge is blending stiff and soft components in a hierarchical architecture spanning several orders of length scale.[8] This design trait of biominerals turns their mechanical response into a complex function of contributions from different length scales. Cortical bone, for instance, contains an array of intrinsic and extrinsic toughening mechanisms that synergistically contribute to its overall fracture toughness.[7,9]

Biological materials often feature structural or compositional gradients imparting a progressive change in specific material properties.[10] For example, such gradients were found, *inter alia*, to increase contact damage resistance[4,5,10–15] or to confer interfacial strengthening and toughening.[10,16,17] Stomatopods are an outstanding example of the efficacy of this design strategy.[18] These mantis shrimps are noted for their aggressive style of hunting since they employ their dactyl club as an ultrafast hammer to disrupt their preys' shells and exoskeletons. Multiple compositional gradients have been identified, providing the club with the required impact and damage tolerance.[4,5,13,15,19]

The black-lipped pearl oyster *Pinctada margaritifera* (Pterioida order) is a typical prey for stomatopods;[20] it is a marine bivalve endemic to intertidal and subtidal habitats of coral reefs. Its shell consists of an inner nacreous layer and an outer layer composed of calcitic columnar prisms. The oyster *P. margaritifera* is not the only specimen with a nacroprismatic shell design; a considerable number of mollusks feature a comparable shell organization. Nacroprismatic bivalves are found in various habitats and feature subtle differences in their shell structures while exhibiting remarkable mechanical properties. For instance, the biogenic and prismatic calcite layers of the bivalves *Atrina rigida* and *Pinna nobilis* are 50–70 % harder than geologic calcite.[21,22] This enhancement originates primarily from hindered dislocation motion, induced by effects such as solid-solution strengthening by $Mg^{2+}$ incorporation[23,24] or by composite strengthening due to incorporated

bio(macro)molecules.[25–31] Remarkably, the typical columnar shape of the tightly tessellating calcite prisms is not crystallographically controlled; it is rather the result of a pseudomorphic phase transformation of a transient amorphous calcium carbonate precursor.[29,32,33] It has been further suggested that the morphogenesis of nacroprismatic shells is guided by thermodynamic constraints such as directional solidification[34] or grain growth.[34,35]

Biogenic calcite prisms that appear to be nearly single-crystalline have probably attracted the most attention. However, only a few species generate such uniform crystalline prisms, such as *P. nobilis* or *A. rigida* with their [*c*]-axes parallel to the growth axis of the prisms and, thus, perpendicular to the shell surface.[36,37] The calcite prisms of other species are crystallographically more complex, with various mapping experiments demonstrating their polycrystallinity. Thus, these prisms are composed of crystallographically iso-oriented subdomains (or crystallites) with distinct misorientation relative to their neighbors.[38–42] For *P. margaritifera*, Checa et al. showed that prisms initially consist of only one crystallographic subdomain, which gradually undergoes crystal lattice tilting and eventually splits into independent subdomains.[43] These subdomains progressively align their [*a*]-axis with those of their neighbors' subdomains, a process that was explained in terms of growth competition between individual crystallites within a single prism.[44] In the light of these contributions, these crystallographic peculiarities have been discussed as an incidental outcome of a biologically controlled crystal growth process, overlooking any further advantage for the biomineralizing organism.[45,46]

However, the strong evolutionary pressures caused by the biological arms race in a given habitat imply that evolutionarily-conserved motifs are likely linked to (unforeseen) functions that increase survival rates.[6] A comparative study on calcitic columnar prisms of seven mollusk shells speculated about a biological function of crystal lattice tilting beyond mere biomineral growth. In that study, the calcite prisms of the two oysters *Pinctada fucata* and *P. margaritifera* were noted for having the most pronounced crystal lattice tilting and the highest hardness values.[47] It was concluded that tilting crystal lattices "is the structural property that confers increased hardness to the prisms of the *Pinctada* species". However, both bivalves also have the smallest crystallite size among those tested.[47] Ultimately, the underlying hardening mechanism remained unsettled as it could equally arise from a Hall-Petch effect, as shown in avian eggshells,[48] or from the further impediment of dislocation motion due to increased crystallographic inhomogeneities.[7,21,49]

In this contribution, we assess the function of the peculiar crystallographic design of the prismatic calcite layer of *P. margaritifera*, through electron backscatter diffraction (EBSD) and a range of complimentary techniques, including indentation experiments and stress-field simulations. We demonstrate for the first time a systematic crystallographic texture gradient across the entire prismatic calcite layer. The gradient is generated by two complementary mechanisms: *(i)* a cross-over in the relative proportions of prisms with different preferential orientation and *(ii)* crystal-lattice tilting in individual prisms. The coaction of these two mechanisms contributes to a less obvious mechanism, *(iii)* gradual ordering of the entire prismatic layer by transition from an initially non- or weakly textured material to a strongly textured material. Since Young's modulus depends on crystal orientation, the texture gradient transforms the prismatic shell layer into a functionally graded material. Finite-element models predict, and indentation experiments demonstrate, that the observed gradient in Young's modulus improves the biomineral's resistance to wear and contact damage. The crystallographic texture gradient imparts stress delocalization, which distinctly improves the shell's fracture toughness, outperforming non-graded counterparts and, thus, increasing the odds for survival.

## 2. Results & Discussion

We employed a range of methods to investigate the outer calcitic layer of the bivalve shell *Pinctada margaritifera* (Figure S1), which consists only of calcite columns separated by a periprismatic organic matrix (**Figure 1**a–b). Prisms are composed of crystallites, which form crystallographic subdomains if their mutual misorientation is small.[43] As shell growth proceeds from the exterior to the interior by layer-by-layer deposition, the shell records a time series progressing from the outside to the inside of the shell (top to bottom in Figure 1a and 1c-f).[45]

## *2.1. Electron backscatter diffraction (EBSD) analyses reveal crystallographic texture gradients*

Electron backscatter diffraction (EBSD) maps of the outer calcitic layer revealed systematic crystallographic texture gradients across its entire thickness, which arise from independent crystallographic changes across length scales, from the crystallite level up to the level of the entire prismatic layer.

The EBSD map presented in Figure 1c covers the complete cross-section of the prismatic layer between the exterior (top) and interior (bottom) boundaries. The crystallographic orientations of the prisms define two end-member groups, with some prisms being transitional between these end-members (Figure 1d–f). Many prisms near the exterior of the shell, Group 1, have [$c$]-axes typically at less than ~50° to the growth direction (Figure 1d). However, the largest area fraction consists of prisms, Group 2, with [$c$]-axes lying approximately within the plane of the shell (Figure 1e). Our classification is in line with recent studies reporting on an unexpected growth direction of late prisms in *P. margaritifera* perpendicular to the [$c$]-axis, i.e., of Group 2 prisms in our terminology.[50]

A smaller set of prisms, termed 'transitional', have [$c$]-axes that start in similar orientations to those of Group 1. However, during progressive growth, the [$c$]-axes of their intraprismatic crystallites rotate by up to 68° to lie approximately within the plane of the shell (Figure 1f); they represent a transmutation of Group 1 prisms into Group 2 prisms. All prisms contain significant orientation gradients, with [*c*]-axis orientations typically varying by several tens of degrees within a single prism, as illustrated in detail in **Figure 2**.

In all groups, the misorientation axes of domain boundaries with misorientation angles in the range 2–10° are parallel to either an <*a*>-axis or {*m*}-pole, i.e., {11$\bar{2}$0} or {10$\bar{1}$0}. However, these misorientation axes have different spatial orientations in each group of prisms (due to their different crystal orientations), leading to differing evolution of lattice orientation during prism growth (Figure S2). The axes of lattice rotation in Group 1 prisms are typically oblique to both the plane of the shell and to the growth direction (Figure 2a and S2). Thus, the crystal lattice tilting in Group 1 prisms rotates the [*c*]-axis towards the plane of the shell during progressive growth, but their growth terminates before their [c]-axes reach the plane of the shell. In Group 2 prisms, out of the three symmetrical equivalents of both {*m*}-poles and/or <*a*>-axes, rotation commonly occurs around the variant parallel to the growth direction (Figure 2b). This constrains the [*c*]-axes of Group 2 prisms to remain within the plane of the shell during growth. In transitional prisms, the [*c*]-axes become aligned parallel to the plane of the shell and remain so during extensive further growth (Figure 2c). Consequently, in all groups [*c*]-axes either rotate toward and/or remain within the plane of the shell during progressive growth.

Overall, we observe a texture gradient due to two obvious mechanisms: *(i)* changes in the relative proportions of prisms from Groups 1 and 2; and *(ii)* crystal lattice tilting, i.e., systematic crystallite-to-crystallite rotations (thus of subdomains within prisms), progressively aligning the <*a*> and {*m*} directions with the growth direction and the [*c*]-direction with the plane of the shell.

The first mechanism can be easily explained by Grigor′ev crystal-growth competition on the prism level, in which crystals with their fastest growth axis oriented approximately perpendicular to the substrate outcompete others and dominate the late stage of prism growth.[51] The crystallographic orientation of the prisms and their geometrical evolution indeed indicate that prism growth rates are crystallographically controlled and that <*a*> and/or {*m*} are the fastest growing directions in this mineralization system. For calcite, growth along the [*c*]-axis is commonly fastest during growth from solution. However, under diffusion-limited conditions, such as by the action of face-specific growth inhibitors or under spherulitic and pseudomorphic growth conditions, other growth directions may take over.[52]

The second mechanism, crystal lattice tilting, is well-documented in synthetic crystalline materials, such as calcite films generated via a nonclassical biomimetic crystallization route.[52,53] In such materials, occluded impurities trigger largely uncontrolled autodeformation and non-crystallographic branching. Specialized intracrystalline biomineralization proteins possibly trigger the misorientations in *P. margaritifera* by comparable, but precisely controlled, mechanisms.[42,43]

These two mechanisms contribute to a third and rather concealed mechanism: *(iii)* gradual crystallographic ordering. The combination of mechansisms *(i)* and *(ii)* result in not only a change in texture but also an increase in texture strength. Prisms of Group 2 (Figure 1F) have a stronger texture (evident as tighter clusters in the pole figures) than prisms of Group 1 (Figure 1D). This effect can be quantified using the J-index,[54] which provides a measure of texture strength. Group 1 has a J-index of 5.5 whereas Group 2 has a J-index of 8.9. This contrast in texture strength records a progressive disorder-to-order transition across the shell thickness.

**2.2. Crystallographic gradients modulate mechanical properties and can increase flaw- and contact-damage tolerance**

*2.2.1. Changing crystal textures alter the Young's modulus normal to the shell surface*
The texture gradients inevitably change the mechanical properties of the entire prismatic layer as the constitutive crystallites have anisotropic mechanical properties. We demonstrate this effect using the directionality of Young's modulus in calcite, illustrated in **Figure 3**a, as the anisotropic elastic properties of constituent crystallites control the effective stiffness of the prismatic layer. We employed effective medium theory based on crystal orientation distributions and the anisotropic elastic stiffness tensor for calcite to estimate Young's modulus perpendicular to the shell surface.[55,56] Mechanical properties in this direction are of particular interest as this is a likely direction of impacts to the shell.

In *P. margaritifera*, Young's modulus exhibits systematic gradients across the mapped area, both within individual prisms (Figure 3c), and across the overall microstructure (Figure 3b). Figure 3b demonstrates that, in general, Young's modulus normal to the shell surface increases with progressive growth. The initially low and heterogeneous Young's modulus is gradually transformed into a high and homogeneous Young's modulus during the first ~100 µm of growth of the prismatic layer (Figure 3b–c and Figure S3).

*2.2.2. The exterior of the calcitic layer optimizes the shell's resistance against wear and crack initiation*

The variable elastic properties necessarily alter the macroscale mechanical performance of the prismatic layer, which are crucial for the mechanical resistance of the protective shell.

The crystallographic layout of the prismatic layer is optimized for greatest wear resistance at the exterior (Figure 3b). We performed nanoindentation parallel to the growth direction at a range of levels within the prismatic layer, which revealed the hardness to be approximately constant at 4.28 ± 0.35 GPa. This value agrees well with the literature and is comparable to other nacroprismatic bivalves such as *P. nobilis*.[22,57] As wear resistance scales with the relation $H^m/E^n$ ($m \geq n \geq 1$), where $H$ is hardness and $E$ is elastic modulus,[58] wear resistance of the prismatic layer is highest at the exterior where the material is most compliant.

We experimentally characterized intraprismatic crack initiation by nanoindentation, which probes length-scales up to a few micrometers, over which Young's modulus is approximately constant and thus is not affected by the gradient.[67–69] Displacement bursts in the continuous load-displacement curves coincided with the appearance of visible cracks and therefore resulted from crack initiation (Figure S4). The probability distributions of crack initiation as a function of indenter force demonstrate that the compliant prisms of Group 1, which dominate the shell exterior, consistently require greater indenter loads to initiate fracture than the stiffer prisms of Group 2 (Figure S4e).

*2.2.3. The internal texture gradients of the calcitic layer improve the shell's stress distribution and flaw-tolerance*

The progressively changing crystal texture, which imparts a change in stiffness normal to the shell surface, has potential to transform the prismatic layer into a functionally graded material that is toughened by reducing the stress at a crack tip. Theoretical studies found that elastic gradients can confer distinct flaw-tolerance through their impact on stress distribution, albeit this effect is sensitive to other material characteristics, such as the Poisson ratio.[59–61] Experimentally, Suresh and co-workers have impressively evidenced improved contact-damage tolerance for the case of graded silicon nitrides.[62–64] Such results point to an additional biological function of graded texture in biominerals: crystallographic gradients may serve as a toughening motif increasing the contact-damage tolerance of mineralized armatures.

We employed finite-element modeling simulations, which suggest that the observed stiffness gradient remarkably toughens the calcitic layer as it reduces the stress intensity factor distinctly. In these simulations, we studied a surface crack under a tensile load by subjecting a notched strip of material to a pure remote mode I loading. For comparison, materials with homogeneous, inverse graded, and discrete, multi-layered Young's modulus were modeled under identical loading conditions (Figure S5). For each case, the fracture resistance is quantified in terms of the predicted stress intensity factor ($K_I$) as a function of crack length, plotted in Figure 4a–b. By exhibiting the lowest stress intensity factor, the graded material outperforms all control materials. The graded material experiences stress intensity factors up to 20 % lower than the homogenous material, improving its fracture resistance. In contrast, the opposite is true when the gradient is reversed. By distributing the stress widely across the bulk, the gradient meliorates the damage tolerance as it diminishes the stress intensity at the crack tip, illustrated in Figure 4c.

We undertook further indentation experiments that corroborated our theoretical predictions. Crack formation caused by indentation experiments has been shown to occur under mode I loading, making these experiments comparable to our simulations.[65,66]

To probe the impact of the observed stiffness gradient on the fracture toughness, we conducted comparative Vickers microhardness tests. In the case of brittle solids, these tests probe depth ranges of approximately 75–100 µm, the scale over which Young's modulus of the oyster shell varies on the order of 25 %. As controls, we used both geological calcite and the prismatic calcite layer of the Mediterranean bivalve shell *Pinna nobilis*. In contrast to *P. margaritifera*, the single-crystalline prisms of *P. nobilis* are composed of co-oriented crystallites with [*c*]-axes normal to the plane of the shell, see Figure S6.[32,70] Thus, the prismatic layer of *P. nobilis* can be seen as a non-graded analog to the crystallographically graded prismatic shell layer of *P. margaritifera* with a crystallographic orientation similar to Group 1 prisms of *P. margaritifera*.

Vickers indentation under identical loading conditions resulted in extended fracture, shattering, and disaggregation of both geological calcite and *P. nobilis* prisms in and around the indents, illustrated in Figure 4d. In stark contrast, *P. margaritifera* endured the indentation with only minimal fracture. The Vickers tests further allowed for quantification of the materials' fracture toughness. Geological calcite exhibited a fracture toughness of 0.34 ± 0.04 MPa·m$^{1/2}$, whereas *P. nobilis* demonstrated a distinctly increased fracture toughness of 1.16 ± 0.49 MPa·m$^{1/2}$, which we attribute to its hierarchical organization (see below). The crystallographically graded prismatic layer of *P. margaritifera* surpasses even *P. nobilis*, with a superior fracture toughness of 3.15 ± 1.86 MPa·m$^{1/2}$.

We assured the general comparability of the hierarchical and ultrastructural organization of prisms of both species by small-angle neutron scattering and anisotropic Scherrer analyses. Both, *P. margaritifera* and *P. nobilis* have been repeatedly reported to feature a similar ultrastructure with nanograins of a few tens of nanometer in size.[27,32,43,70,71] We probed the sub-prismatic crystallite morphology by anisotropic Scherrer analyses,[72] which provided spatial dimensions of the coherently scattering domains. In the case of geological calcite, the crystallite sizes were beyond the range of this approach (> 1000 nm), which demonstrates that this control calcite was well-crystalline. In contrast, we found that the crystallite morphologies in prisms of both bivalves feature a notable and comparable asymmetric layout. Their coherently scattering domains largely stretch out in the (*hk0*) directions (> 1000 nm), but they are confined to some few hundreds of nanometers in the (*001*) direction. Thus, the crystalline domains of *P. nobilis* and *P. margaritifera* prisms have both a prolate shape with comparable aspect ratio and lateral dimensions of 200 ± 50 nm and 450 ± 200 nm, respectively. It should be noted that recent ptychography studies demonstrated that crystalline coherence can also expand across a couple of granules.[42]

We additionally performed small-angle neutron scattering (SANS) both in the small- and very-small angle regime, which confirmed that the hierarchical organization of the prisms is comparable, especially on the sub-micron length scale. Both types of calcite prisms are fractal materials characterized by a Porod exponent of 3, which suggests a three-dimensional network of particles.[73,74] This inference is consistent with the observation that bivalve shell growth can proceed via the aggregation of nanosized particles.[75,76] A multi-level Beaucage fit reveals further structural elements across scales, thus a hierarchical order of building units, see Figure S7.[77] The powdered calcite prisms of *P. nobilis* are composed of three different units with diameters of approximately 0.55 µm, 45 nm, and 1.4 nm, respectively. In the case of *P. margaritifera*, again, three structural units are found, with similar characteristic gyration radii of 0.87 µm, 40.8 nm, and 2.2 nm. We assume that the structural units with a diameter of 0.55 and 0.87 µm reflect the particle sizes of the powdered samples as no larger units are detected. The next structural elements with an approximate diameter of about 40 nm fit well with the reported size of the mineral nanogranules,[27,32,43,70,71] whose self-assembly probably drives shell formation in these bivalves.[75,76] The smallest component identified by SANS is in the low nanometer range. Their diameter is consistent with reports of the intergranular organic matrix, which can be described as a holey organic envelope of the nanogranules or, alternatively, as an interstitial extended network with a thickness of 1–2 nm.[26,32,43,78] We also determined the mass fraction of this intracrystalline organic matrix by thermogravimetric analysis (TGA) and found in both species a similar amount of about 0.5–1 %. The results obtained from two different scattering methods and TGA validate our assumption that the prismatic layer of *P. nobilis* represents a comparable but non-graded counterpart to the crystallographically graded prismatic layer of *P. margaritifera*.

## 3. Conclusions

Our results demonstrate that, concealed by the complex crystallographic texture of its shell, the black-lipped pearl oyster bears a hidden functional gradient that optimizes the molluscan body armor. The gradient remarkably toughens the shells against catastrophic failure by spreading the stress field into the bulk, increasing fracture toughness, and simultaneously optimizing the shell's resistance against wear and crack initiation at its surface.

Using the example of *P. margeritifera*, we demonstrate a design principle of biominerals based on graded material properties. We reveal for the first time that it is a gradient in average crystallographic orientations (crystallographic texture), which effectively hones the mechanical properties of this biomineral. The present case of *P. margaritifera* exemplifies two obvious mechanisms, which both give rise to a gradual change in texture: *(i)* crystal-lattice tilting within prisms and *(ii)* a ratio cross-over of prisms with different preferred crystal orientations. A third mechanism is also present, although in a rather concealed fashion: *(iii)* gradual ordering, i.e., a disorder-to-order transition from weak to strong texture.

Our findings establish that not only species with unidirectionally and near-single-crystalline bioarmor employ crystallographic control to optimize function. The functionally beneficial change of crystallite orientation during growth generates materials with complex and apparently unsystematic texture which stand out with excellent and complex mechanical properties.

The key to increased survival rates by increased contact-damage resilience lies in the generation of a stiffness gradient, which exposes the most compliant material towards the direction of attack. Such stiffness gradients are potentially widespread in protective biomaterials, and we expect that also other design principles, such as amorphous-to-crystalline or increasing ratio of inorganic/organic blends, will also confer increased contact-damage resistance. Thus, we anticipate that a wealth of seemingly poorly ordered biogenic materials (e.g., in Unionidae[34,79]) use comparable concepts to generate functionally graded materials. Complex crystallography of biominerals should, therefore, be perceived as evolutionarily-conserved motifs rather than a mere nonfunctional outcome of an incipiently deficient biological control over crystal growth.

Synthetic materials or ceramics have not yet exploited the directionality of elastic properties in crystalline material and, thus, the black-lipped pearl oyster can serve as a blueprint for future bioinspired functional materials. This elegant concept applies to a wide range of material classes with anisotropic elasticity and thus may generate strong impetus for new design approaches towards architectured materials and novel non-brittle ceramics.

**Experimental Section**

*Electron backscatter diffraction analysis*: Orientation data were processed and plotted using Oxford Instruments HKLChannel5 software. Misindexed pixels with > 10° misorientation from each of their eight neighbors were removed, and unindexed pixels with at least six neighbors within the same prism were filled with the average orientation of their neighbors. Analysis of elastic properties, such as Young's modulus, was performed using the MTEX 4.5 toolbox for MATLAB®, following the approach of Mainprice *et al.* and using the elastic stiffness tensor for calcite of Chen *et al.*[55,56] This approach was preferred over common experimental techniques for quantifying variations in heterogeneous mechanical properties over micrometer length-scales, such as nanoindentation,[21] which also can measure an effective elastic modulus. However, for highly anisotropic materials such as calcite, which exhibits auxeticity,[80] it is not appropriate to use conventional equations based on isotropic elasticity to derive Young's modulus from effective indentation moduli, as they do not account for variation in Poisson's ratio.[21] The method that we employed to predict elastic properties is widely used and tested in the geological sciences,[55,81,82] and previous work suggests that the organic content of biogenic calcite prisms does not cause its elastic properties to deviate significantly from those of geological calcite.[21]

*XRD and Anisotropic Scherrer Analyses*: X-ray diffraction analysis was performed at a D8 diffractometer (Bruker AXS, Karlsruhe, Germany) equipped with a 9-fold sample changer. Specimens were prepared by a front-loading method in quadruplicate. The following measurement parameters were applied: range 10–90° 2θ; step size 0.011° 2θ, integration time 0.4 s; radiation: Cu K$_\alpha$; generator settings: 40 mA, 40 kV; divergence slit: 0.3°. Rietveld refinement was performed by means of the software TOPAS V5 (Bruker AXS, Karlsruhe, Germany. For the refinement of calcite, a hkl phase with the space group and lattice parameters of the calcite structure ICSD #80869 by Maslen et al.[83] was used to obtain an optimum fit. The lattice parameters a and c were refined. For the refinement of crystallite (coherent scattering domain) sizes, an anisotropic crystallinity model was applied, in which the shapes of the crystallites were simulated by suitable geometric models, following the approach of Ectors et al.[72]. A triaxial ellipsoid model was chosen and its dimension was defined by three radii *rx*, *ry*, and *rz*. The radius in *z*-direction, *rz*, is parallel to the crystallographic [*c*]-axis of the trigonal calcite crystallites, and rx is parallel to the [*a*]-axis. Due to the crystallographic requirements of the trigonal system *rx* and *ry* are equal.

***Thermogravimetric analysis:*** Prisms were separated by bleaching, which disintegrated the interprismatic organics and the periostracum while keeping the mineral prisms intact. The prismatic shell layers were cut into small pieces, which were immersed in sodium hypochlorite solution (0.25 g of active chlorine/100 ml of water) for three days. After filtration, the prisms were collected by filtration and rinsed with water. Samples were powdered and subjected to thermogravimetry analysis under nitrogen atmosphere at a heating rate of 5 K min$^{-1}$ (TA Instruments, TGA Q5000).

***Small and ultra-small angle neutron scattering (SANS, VSANS)*:** Small-Angle Neutron Scattering (SANS) experiments were performed on KWS-2 and KWS-3 instruments operated by Jülich Centre for Neutron Science (JCNS) at the Heinz Maier-Leibnitz Zentrum (MLZ) in Garching, Germany.[84,85] With the focusing SANS instrument KWS-3, scattering vectors *Q* from 0.0001 to 0.35 Å$^{-1}$ are accessible, at sample-to-detector distances of 9.2, 1.25, 0.25 and 0.05 m. This resolution was reached by a toroidal mirror with focus-to-focus distance 22 m, entrance aperture 2x2 mm$^2$, wavelength $\lambda$ = 12.8 Å ($\Delta\lambda/\lambda$ = 17), and two-dimensional position-sensitive scintillation detector with diameter 9 cm and pixel size 0.32 mm. To improve scattering statistics above $Q$ = 0.02 Å$^{-1}$ experiments were also carried out at the classical pinhole SANS instrument KWS-2 at sample-to-detector distances of 2 and 8 m (corresponding collimation length of 4 and 8 m), and wavelength of $\lambda$ = 5 Å ($\Delta\lambda/\lambda$ = 10%). Within the configurations mentioned above, KWS-2 covers a $Q$-range from 0.01 to 0.45 Å$^{-1}$. Powders of separated prisms were placed into a demountable quartz cell with a path length of 0.1 mm. The data reduction, analysis, background subtraction and fitting were performed using the software QtiKWS (V. Pipich, 2019, http://www.qtikws.de). A multi-level Beaucage fit was applied.[86]

*Mechanical Testing*: To determine hardness and fracture toughness of the prisms, nano- and microindentation studies were performed on transverse cross-sections of the prismatic layer. Prior to indentation the sections were polished using a Multiprep lapping unit (Allied technologies) with 0.5 µm diamond lapping films. The nanoindentation experiments were performed using a nanoindenter XP (Keysight Technologies Inc., USA) and a Berkovich diamond tip with a maximum penetration depth of 200 nm. The continuous stiffness measurement (CSM) technique, with a frequency of 45 Hz and a harmonic displacement amplitude of 2 nm, was applied to characterize the evolution of the hardness and effective elastic modulus as a function of penetration depth.[87–89] The measurements were started after reaching a stable thermal drift-rate less than 0.05 nm/s. To further account for thermal drift effects, the force was held constant at the end of the unloading step at 10 % of maximum load for 60 s and the displacement data were corrected based on the measured drift rate. Crack-initiation studies were undertaken on at least 12 stiff and 12 compliant prisms. Nanoindentation measurements were performed in a load-controlled manner up to a maximum force of 15 mN using a diamond Berkovich indenter tip. Pop-ins caused by crack initiation were defined as a burst in the load-displacement curve of greater than 3 nm based on analysis of indents from tests with and without pop-in events. After indentation, samples were investigated with a Dimension 3100 atomic force microscope (Bruker Corporation, USA) operating in contact mode. Standard silicon nitride contact tips were used with a nominal tip radius of 20 nm. Image acquisition was performed at a scan-rate of 0.5 Hz and a resolution of 512x512 pixels.

Vickers micro-indentation studies were performed on transverse cross-sections of *P. margaritifera* and *P. nobilis*, and geological calcite prepared using the same protocol as for nanoindentation studies. Geological calcite single crystals were indented parallel to the [*c*]-axis. The crystallographic orientation of the prepared geological calcite sample surface was assessed using EBSD. Samples were indented using a Zwick 3212 indenter (Zwick GmbH, Ulm, Germany) with 1 kg loading with 10 seconds loading time. Samples were subsequently sputter coated with gold and imaged with secondary electrons in a scanning electron microscope. For fracture toughness determination from Vickers indents, cracks were classified as Palmqvist-cracks. Fracture toughness was then determined using Equation 1 (given below), under the assumption that the cracks are approximately elliptical, and that horizontal crack length is therefore also a proxy for vertical crack depth.

$$K_{\mathrm{IC}} = 0.032\, H\sqrt{a} \cdot \sqrt{\frac{E}{H}} \cdot \left(\frac{c}{a}\right)^{-\frac{3}{2}} \qquad (1)$$

*Finite-Element Modelling:* To predict the effect of the gradient in Young's Modulus within the prismatic layer of *P. margaritifera* on the toughness of the material, we constructed finite-element models (FEMs) of a long strip of material with a pre-existing crack loaded in Mode I (Figure S5A–B). This particular condition mimics the condition induced by the indenter when the crack tip is sufficiently far away from the indenter. We modelled four distributions of Young's Modulus indicated in Figure S5C: the graded distribution based on the EBSD-derived distribution in *P. margaritifera* (red line), the inverse graded distribution (blue line), a discrete layered distribution (green line), and a homogeneous distribution (black line). The finite-element computational domain is shown in Figure S5D. The length and width of the domain were 4.8 mm and 0.6 mm, respectively. For the layered material, the domain was divided into 6 regions of width $w_0 = 0.1$ mm with constant Young's modulus. The values of Young's modulus were 79.4 GPa, 91.0 GPa, 92.8 GPa, 88.3 GPa, 86.5 GPa, and 86.4 GPa. For each case, we considered finite-element meshes with seven crack sizes, $a = 1$ μm, 12 μm, 25 μm, 38 μm, 50 μm, 75 μm, 100 μm, 125 μm, and 150 μm. The FEM analysis was performed in the software package Abaqus and we employed 2D plane strain elements (i.e., element CPE4R). We applied a uniform displacement at the top boundary and symmetry boundary conditions at the bottom of the mesh. The mesh was refined at the crack tip, as displayed in Figure S5D. The mesh size around the crack tip was determined by a preliminary convergence study on the values of the J-integral with three different mesh sizes, presented in Figure S5H. Thereby, the element size near the crack tip was determined to be small enough to provide an accurate stress field. To determine the stress intensity factor variation at the crack tip for the various cases, we employed a "modified" J-integral because the conventional J-integral method (also implemented and available in Abaqus) could be path dependent for the heterogeneous case. Conversely, the modified J-integral is independent of the contour path that is chosen. Based on previous works,[90–95] we adopt the following J-Integral:

$$\bar{J} = \sum_A \sum_{p=1}^{N} \left\{ \left[ \sigma_{ij} u_{i,1} q_{1,j} - W q_{1,1} - W_{,1} q_1 \right] \det \frac{\partial x_k}{\partial \zeta_l} \right\}_p w_p \tag{1}$$

where, $w_p$ are the weights of integration, det() is the determinant of Jacobian matrix, $W$ is the strain energy density, $u_{i,1}$ is the derivative in the *x*-direction of the displacements, $q_1$ is a continuous function with value zero on the outer contour and value one on all the other contours except the outer path.[90–95] $N$ is the number of integration points, $A$ is the area of integration contour, and $\sigma_{ij}$ is the components of the Cauchy stress tensor. We note also that $W_{,1} = \frac{dW}{dx} = \frac{dW}{dE(x)} \cdot \frac{dE(x)}{dx}$, E is a function of $x$, and $W = 0.5(\varepsilon_{11}\sigma_{11} + \varepsilon_{22}\sigma_{22} + \gamma_{12}\sigma_{12})$.

Based on the constitutive law for plane strain, we can derive the equation of $\frac{dW}{dE(x)}$ using the chain rule:

$$q_{1,i} = \begin{Bmatrix} \frac{\partial q_1}{\partial x} \\ \frac{\partial q_1}{\partial y} \end{Bmatrix} = \begin{bmatrix} \frac{\partial x}{\partial \xi} & \frac{\partial y}{\partial \xi} \\ \frac{\partial x}{\partial \eta} & \frac{\partial y}{\partial \eta} \end{bmatrix}^{-1} \begin{Bmatrix} \frac{\partial q_1}{\partial \xi} \\ \frac{\partial q_1}{\partial \eta} \end{Bmatrix} \quad (2)$$

where $\xi$ and $\eta$ are orthogonal axes of the standard quadrilateral element, $q_1 = N_m Q_m$, N is the shape function of each element, and Q is the scalar value 1 or 0. Finally, we use equation (3) to calculate stress intensity factor $K_I$.

$$K_I = \sqrt{\frac{\overline{J} \cdot E_{tip}}{1-v^2}} \quad (3)$$

While the conventional J-integral is expected to be path dependent, previous works [10-11] have demonstrated that under very specific conditions (e.g., for very small contours within refined mesh close to the crack tip), the modified and conventional J-integral can give the same results. A comparison between the modified and conventional J-integrals is presented in Fig. S5E for graded material and in Fig. S5F for inverse graded material. The values of the modified and conventional J integral close to the crack tip for both materials are equal. Modified J integral values are independent of contours of elements. One element contour includes ring elements around the crack tip, presented in Fig. S5D. As expected, the conventional J-integral begins to give different results for larger contours. A close-up plot of the normal stress distribution along the thickness of the strip of material is presented in Fig. S5G. Normalized stress intensity factors of heterogenous materials $K_{I,het}^*$ were calculated on basis of equation (4), where $K_{I,hom}$ is the non-normalized stress intensity factor of the homogeneous material and $K_{I,het}$ that of the heterogeneous material.

$$K_{\text{I},het}^{*} = \frac{K_{\text{I},hom} - K_{\text{I},het}}{K_{\text{I},hom}} \qquad (4)$$

**Supporting Information**
Supporting Information is available from the author.

**Acknowledgements**

DW and JH contributed equally to this work. SEW acknowledges financial support by an Emmy Noether starting grant by the Deutsche Forschungsgemeinschaft (DFG, German Research Foundation), grant number 251939425. SL was supported by the Bavarian State Ministry of the Environment and Consumer Protection in the framework of the project network *BayBionik*, a collaborative research group (sub-project TUT01UT-73842, grant holder SEW). DW and LNH acknowledge support from the Natural Environment Research Council Grant NE/M000966/1. DW and PZ acknowledge financial support from the Multi-University Research Initiative (AFOSR-FA9550-15-1-0009). BM and PF acknowledge financial support from DFG via the research training school GRK1896: "In-Situ Microscopy with Electrons, X-rays and Scanning Probes" and the 'Center for Nanoanalysis and Electron Microscopy' (CENEM) at FAU. We thank J. Reiser and Dr. J. Kaschta and the Institute of Polymer Materials (LSP) at FAU for their support in TGA measurements.

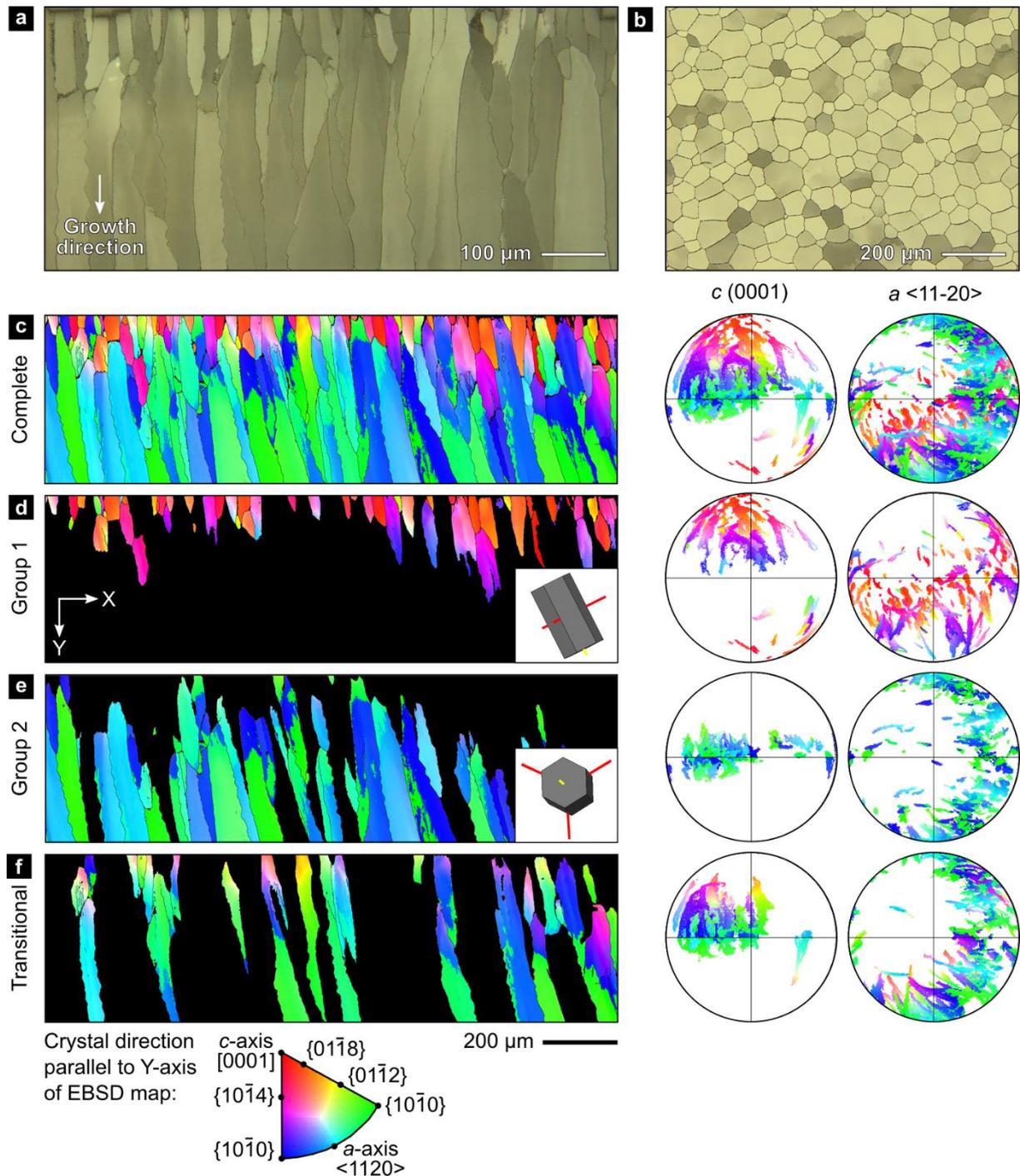

**Figure 1.** Crystallographic orientations in the prismatic layer of *Pinctada margaritifera*.
**a and b,** Polarized reflected light micrographs of sections of the prismatic layer perpendicular and parallel to the shell surface. **c to f,** EBSD maps and corresponding pole figures for the complete dataset and subsets of Group 1 prisms, Group 2 prisms, and Transitional prisms. All maps and pole figures are in the same reference frame and are colored according to the inverse pole figure legend indicating the crystallographic direction aligned with the Y-direction of the map, i.e., perpendicular to the shell surface. The inset images in subfigure d and e indicate typical crystal orientations within each group.

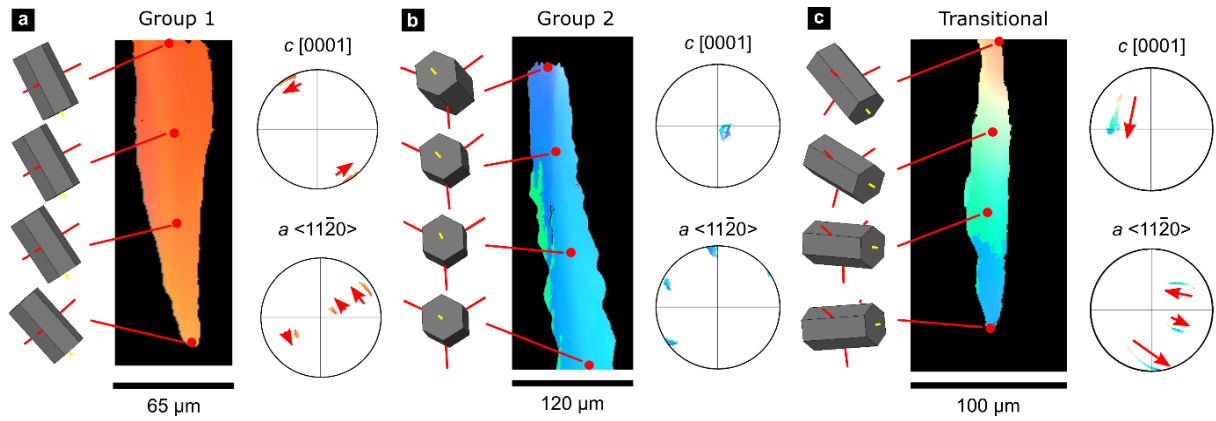

**Figure 2.** Gradients in crystallographic orientation within calcite prisms of *P. margaritifera*. Maps and pole figures of Group 1, Group 2, and transitional prisms colored according to the inverse pole figure for the Y-direction, inset in Figure 1c.

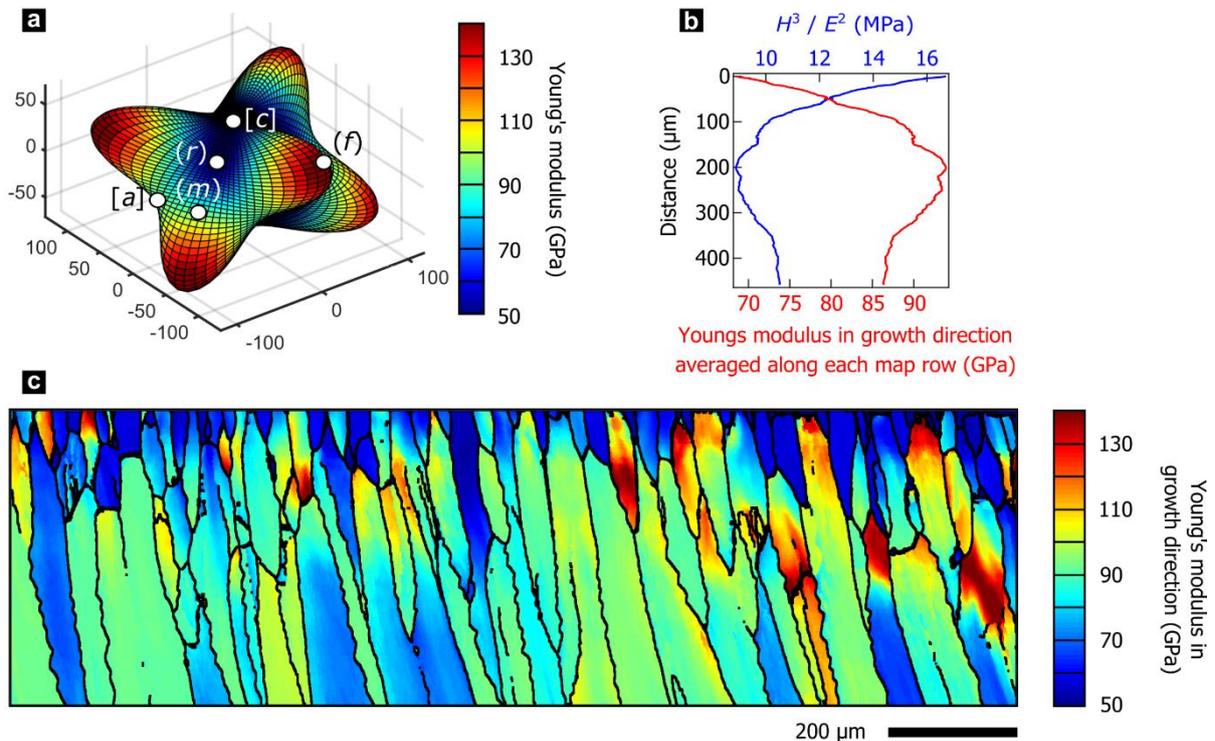

**Figure 3. a,** Three-dimensional visualization of the anisotropic Young's modulus in a calcite single crystal. **b,** Young's modulus in Y-direction averaged per line and $H^3/E^2$, a measure of wear resistance, as a function of distance from the shell exterior. **c,** Calculated Young's modulus in the Y-direction of the map, i.e., perpendicular to the shell's surface and, thus, parallel to the growth direction.

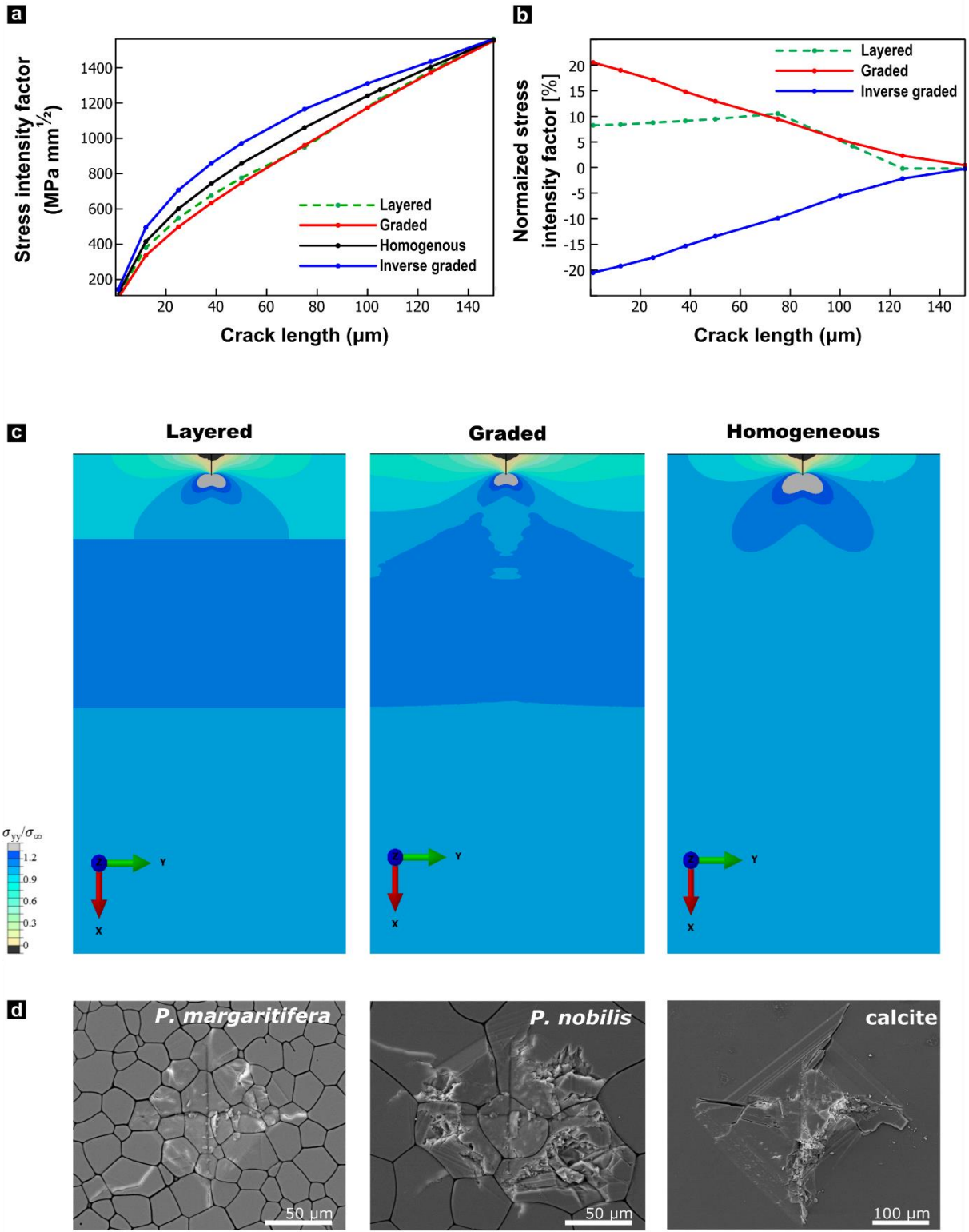

**Figure 4.** Mechanical impacts of the functional gradient in *P. margaritifera* predicted by finite-element modelling and tested by Vickers microindentation. **a**, Stress intensity factor at the crack tip as a function of crack length. **b**, Stress intensity factor of graded, inverse graded, and layered materials normalized with respect to the homogeneous material, as a function of crack length. **c**, Spatial distributions of the $\sigma_{yy}$ component of stress normalized by the applied boundary stress ($\sigma_\infty$) around a 25 µm crack. **d**, Optical micrographs of comparative Vickers microhardness tests on *P. margaritifera*, *P. nobilis*, and geological calcite.

Biominerals are often highly praised for their exceptional control of crystallinity as some organisms generate nearly single-crystalline biocrystals. However, biominerals overwhelmingly show intricate and gradually changing crystal textures, the biological functions of which are not understood. The case of *Pinctada margaritifera* presented herein demonstrates that texture gradients represent functional graded materials that endure due to toughening mechanisms, such as stress delocalization.

**Keywords** Bioceramics

D. Wallis, J. Harris, C. F. Böhm, D. Wang, K. Hurle, P. Zavattieri, P. Feldner, B. Merle, L. N. Hansen, F. Marin, and S. E. Wolf*

**Biominerals with Texture Gradients are Functionally Graded Bioceramics Toughened by Stress Delocalization**

**ToC figure** ((Please choose one size: 55 mm broad × 50 mm high **or** 110 mm broad × 20 mm high. Please do not use any other dimensions))

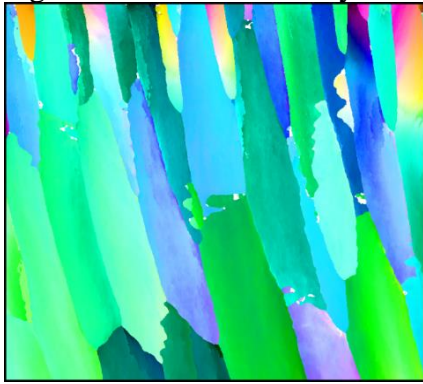

**Biominerals with Texture Gradients are Functionally Graded Bioceramics Toughened by Stress Delocalization**

*David Wallis, Joe Harris, Corinna F. Böhm, Di Wang, Pablo Zavattieri, Patrick Feldner, Benoit Merle, Vitaliy Pipich, Katrin Hurle, Lars N. Hansen, Frédéric Marin, and Stephan E. Wolf\**

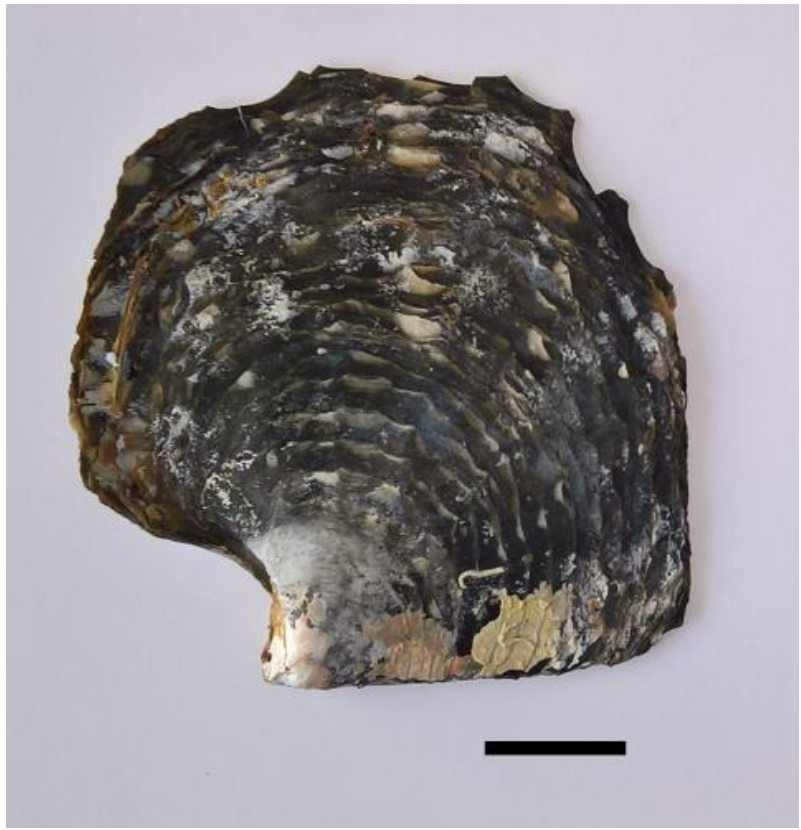

**Figure S1. Macrophotograph of *Pinctada margeritifera*.** Scale bar: 4 cm.

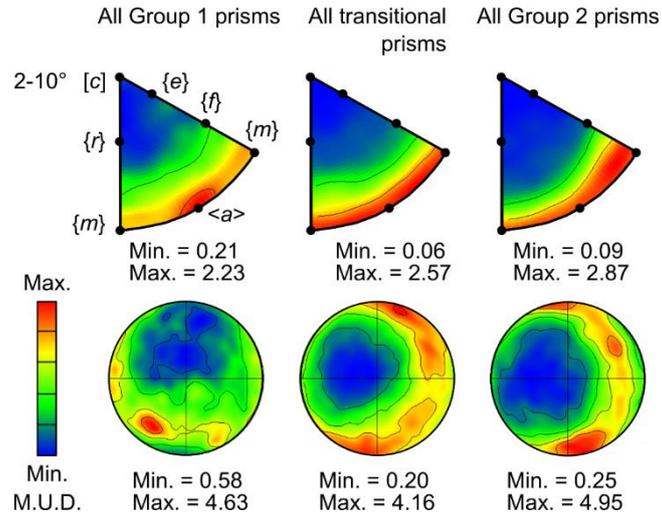

**Figure S2. Axes of 2–10° lattice rotations within Group 1, transitional, and Group 2 prisms.**
Plots are presented in both the crystal (top) and sample (bottom, lower-hemisphere) reference frames. In all groups, rotation axes are approximately perpendicular to the [*c*]-axis and exhibit some bias towards <*a*>-axes in Group 1, consistent with dispersions of crystal axes in pole figures in Figure 2. Rotation axes in Group 1 prisms are predominantly oblique to both the growth direction and the plane of the shell, whereas those in Group 2 prisms are predominantly parallel to the growth direction. Contours are multiples of uniform distribution (M.U.D.).

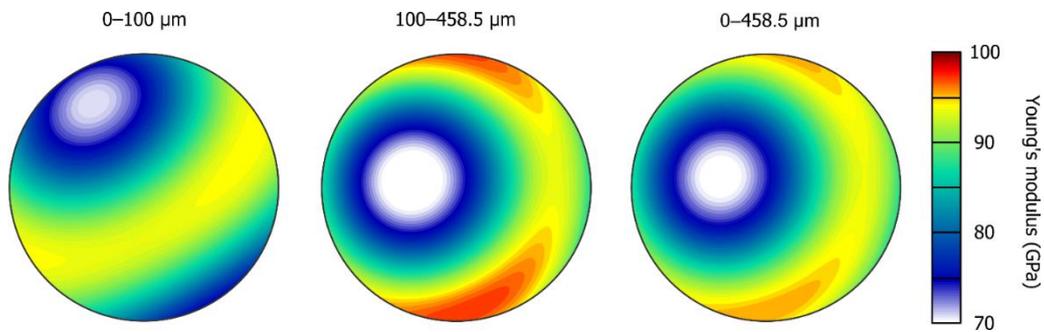

**Figure S3. Anisotropy in Young's modulus**
Stereoplots of Young's modulus for different portions of the prismatic layer, calculated from the Hill average of the Voigt and Reuss bounds on the elastic stiffness tensor[55,81]. Distances correspond to the Y-axis in Figure 1. All distributions are highly anisotropic. In the outer 100 μm of the prismatic layer, the maximum Young's modulus is oriented approximately 60° oblique to the surface normal of the shell, yielding a Young's modulus of ~80 GPa perpendicular to the shell surface. In the remainder of the prismatic layer, the maximum Young's modulus aligns with and remains parallel with the surface normal of the shell, yielding a Young's modulus of ~95 GPa.

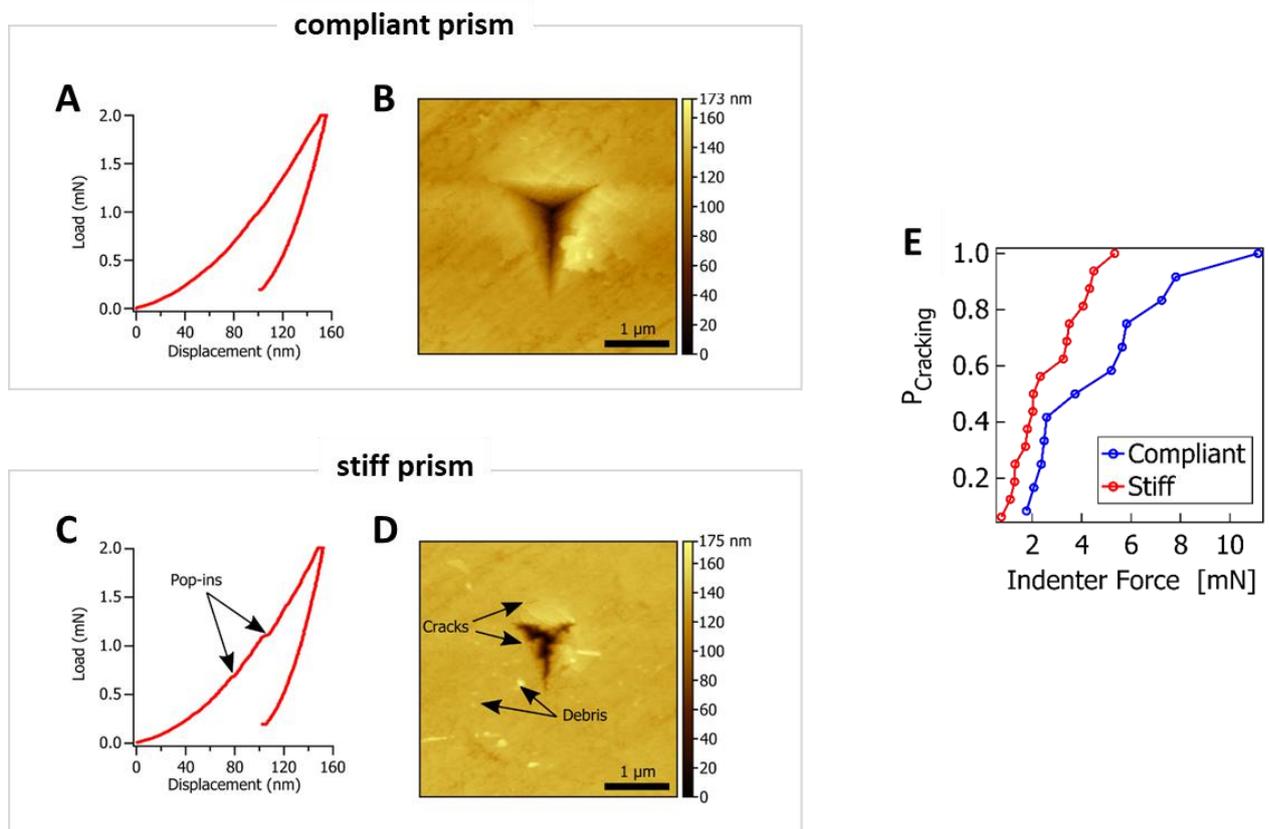

**Figure S4. Nanoindentation analysis of compliant and stiff prisms of *P. margaritifera*.**
(**A**) Load-displacement curve recorded from a test on a compliant prism close to the shell exterior; no pop-ins are visible. (**B**) AFM image of the indent corresponding to the load-displacement curve in A; no cracks are visible. (**C**) Load-displacement curve recorded from a test on a stiff prism close to the shell interior; pop-ins are clearly visible. (**D**) AFM image of the indent corresponding to the load-displacement curve in C. Multiple cracks are present at the edges of the indent and surface debris is present due to cracking. (**E**) Probability of crack initiation as a function of load for stiff and compliant prisms measured using nanoindentation.

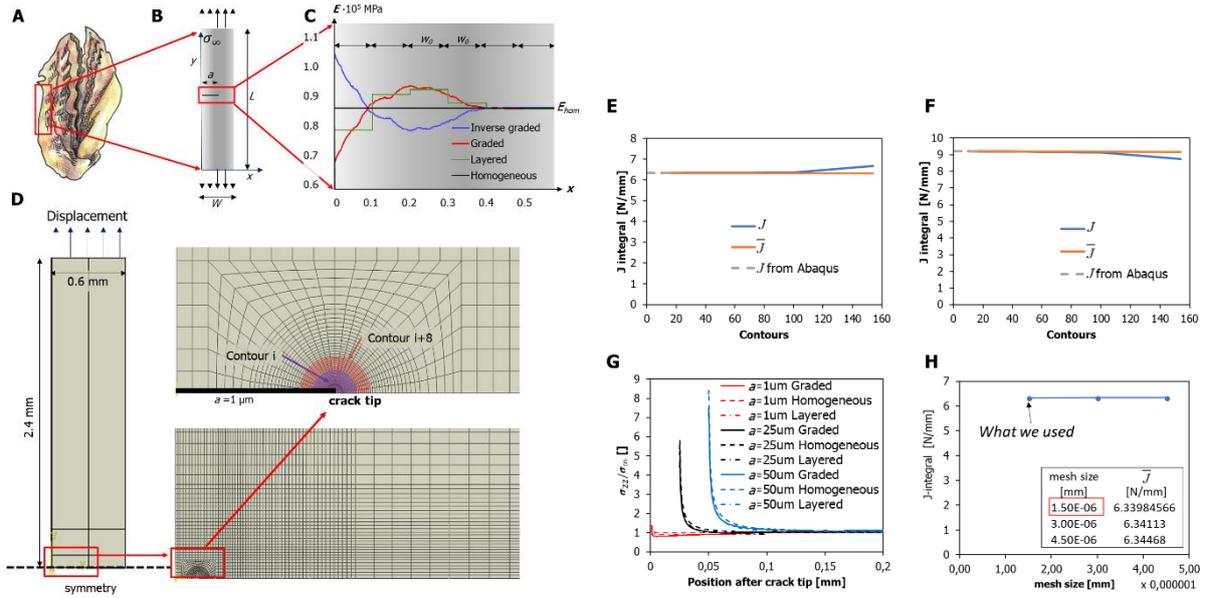

**Figure S5. Details on the finite-element modelling.**
(**A, B**) FEM model description. Original locus of the material strip and crack tip subjected to uniaxial stress conditions imposing a Mode I loading condition at the pre-existing crack.
(**C**) Young's modulus distributions considered in the model: graded, inverse graded, layered, and homogeneous distributions. (**D**) Mesh details with loading and boundary conditions (example shown with a crack size a = 1 µm). (**E**) Comparison of conventional integral calculation on J-integral values (blue) and "modified" J-integral calculation for graded materials (orange), coupled with the domain integral prediction from ABAQUS. (**F**) Comparison of domain integral calculation on J-integral values (blue) and modified J-integral calculation for inverse graded materials (orange), coupled with the domain integral prediction from ABAQUS. (**G**) Normalized opening stress distribution close to crack tip for three different sizes of crack model. (**H**) Mesh convergence study on model with crack size a=50 µm.

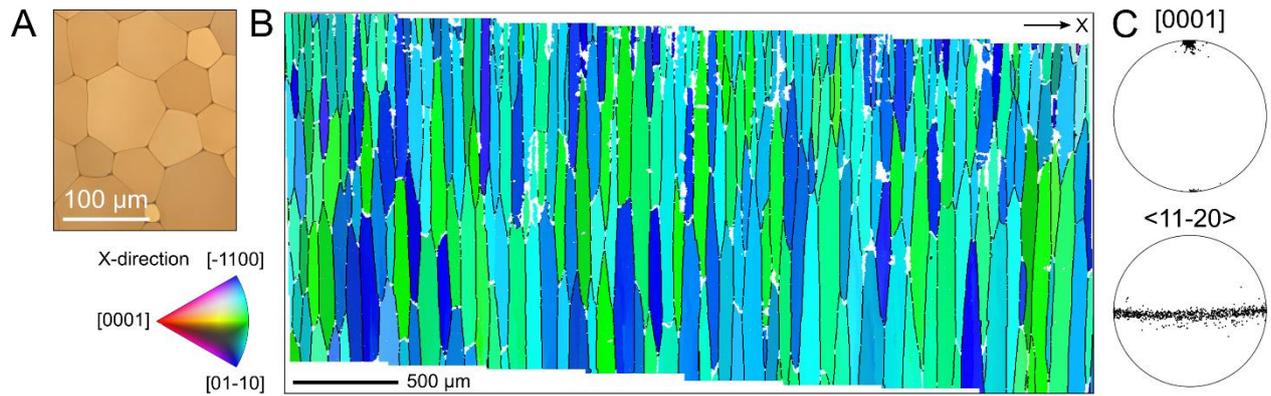

**Figure S6. The prismatic layer of *Pinna nobilis***

**(A)** Reflected light micrograph of a transverse section exhibiting tessellating polygonal prisms. **(B)** Electron backscatter diffraction map of a longitudinal section colour-coded by the inverse pole figure for the crystal direction aligned with the X-axis. No orientation gradients within prisms or systematically across the layer are evident. **(C)** Pole figures of the c[0001] and a<11-20> axes of the prisms in B. [0001] axes are aligned parallel to the growth direction and <11-20> axes are parallel to the plane of the shell.

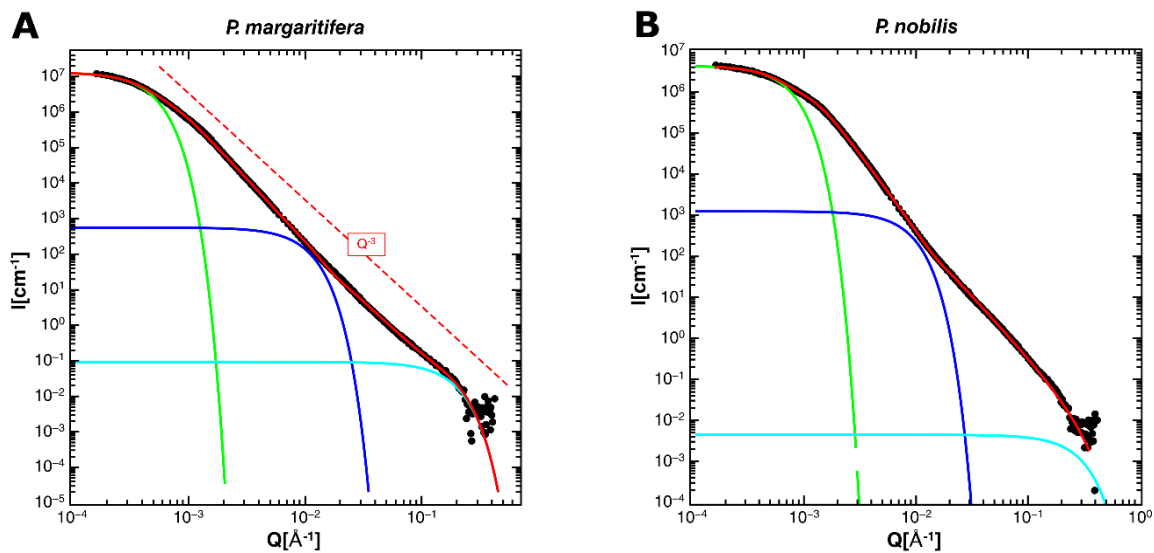

**Figure S7. Small and very-small angle neutron scattering on powdered prisms extracted from nacroprismatic bivalves.**

The experimental data is given by full circles. The prisms of both bivalves, *P. margaritifera* and *P. nobilis* are both characterized as a mass fractal with a Porod exponent of 3. A red dotted line gives a $Q^{-3}$ as a visual guide. Full colored lines give the respective fit based on a multi-level Beaucage form factor model. The calcite prisms of *P. nobilis* are composed of three different units characterized by a radius of gyration of $275 \pm 7$ nm, $22.5 \pm 0.5$ nm, and $0.7 \pm 0.01$ nm, respectively. In the case of *P. margaritifera*, again, three structural units are found, with similar characteristic gyration radii of $435 \pm 10$ nm, $20.4 \pm 1$ nm, and $1.11 \pm 0.03$ nm.